\documentstyle[11pt,paspconf,epsf]{article}

\begin{document}

\vspace*{-25mm}
\noindent
To appear in {\em The Low Surface Brightness Universe}, the proceedings 
of I.A.U. Colloquium No.~171, Cardiff, 6th--10th July, 1998, 
eds. J.~I.~Davies, C.~D.~Impey \& S.~Phillipps, publ. Astronomical 
Society of the Pacific, in press.         
\\[11mm]

\title{Environmental Effects on the Faint End of the Luminosity Function}
\author{S. Phillipps, J.B. Jones}
\affil{Department of Physics, University of Bristol, Bristol, UK}
\author{R.M. Smith}
\affil{Department of Physics and Astronomy, University of Wales, Cardiff, UK}
\author{W.J. Couch, S.P. Driver}
\affil{School of Physics, University of New South Wales, Sydney, Australia}

\begin{abstract}
Recent studies have demonstrated that many galaxy clusters
have luminosity functions (LFs) which are steep at the faint end. However,
it is equally clear that not all clusters have identical LFs. In this
paper we explore whether the variation in LF shape correlates with
other cluster or environmental properties.
\end{abstract}

\section{Introduction}

Much recent work has been devoted to measuring the galaxy luminosity function
(LF) within rich clusters, particularly with regard to the faint end which has
become accessible to detailed study through various technical and 
observational improvements (see the paper by Smith et al. in these 
proceedings).
These studies suggest that the LF becomes steep (Schechter (1976) slope 
$\alpha \leq - 1.5$) in many clusters,
faintwards of about $M_{B} = -17.5$ or $M_{R} \simeq -19$ (for
$H_{0}$ = 50 km s$^{-1}$ Mpc$^{-1}$), where (generally low surface
brightness) dwarfs begin to dominate (e.g. Smith, Driver \&
Phillipps 1997; Trentham 1997a,b). 
Using deep CCD imaging from the Anglo-Australian Telescope,
we have now extended this work (see Driver, 
Couch \& Phillipps 1998), in order to examine the
luminosity distribution in and across a variety of Abell and ACO
clusters. In particular, we were interested in any
possible dependence of the dwarf population (specifically
the ratio of the number of dwarfs to the number
of giants) on cluster type or on position
within the cluster. 

\section{Dwarfs in Rich Clusters}

A number of  
papers (e.g. Driver et al. 1994; Smith et al. 1997; Wilson et al. 1997)
have demonstrated remarkably similar dwarf populations in
a number of morphologically similar, dense rich clusters like (and
including) Coma. This similarity appears not only in the faint end
slope of the LF, around $\alpha = -1.8$, but also in the point at
which the steep slope cuts in, $M_{R} \simeq -19$ (i.e. about $M^{*} + 3.5$). 
The latter implies equal ratios
of dwarf to giant galaxy numbers in the different clusters.

However, there clearly do exist differences between some clusters. 
For example,
several of the clusters in the Driver et al. (1998) sample do not
show a conspicuous turn up at the faint end (see also Lopez-Cruz
et al. 1997 for further examples). Either these clusters
contain completely different types of dwarf galaxy population or, as we
suggest, the turn up occurs at fainter magnitudes. For a composite
giant plus dwarf LF, this is equivalent to a smaller number of dwarfs
relative to giants. 

To simplify the discussion, we will define the dwarf to giant ratio DGR 
as the number of galaxies with $-16.5 \geq M_{R} \geq -19.5$
compared to those with $-19.5 \geq M_{R}$),\\
i.e. $\: \: \: \: DGR = \frac{N(-16.5 > M_{R} > -19.5)}{N(-19.5 > M_{R} > -23.5)}$ . 

The DGR does not have any obvious dependence on cluster richness
(Driver et al. 1998; see also Turner et al. 1993), but
we can also check for variations with morphological characteristics
of the clusters. For giant galaxies, it is well known that a cluster's
structural and population characteristics are well correlated.
For example, dense regular clusters are of early Bautz-Morgan type
(dominated by cD galaxies) and have the highest fractions of giant
ellipticals (Dressler 1980). 
In a similar way, we find that the DGR (i.e. the fraction
of dwarfs) is {\it smallest} in these early Bautz-Morgan type clusters 
(Driver et al. 1998).
Next consider the galaxy density. We can characterise the clusters by their
central (giant) galaxy number densities, for instance the number of 
galaxies brighter
than $M_{R} = -19.5$ within the central 1 Mpc$^{2}$ area. An alternative
would be to use
Dressler's (1980) measure of the average number of near neighbours. 
We then find (solid squares in Figure 1) that the clusters with the least
prominent dwarf populations (low DGRs $\sim 1$)
are just those with the highest projected
galaxy densities (e.g. the Bautz-Morgan Type I-II cluster A3888). 
Previously, Turner et al. (1993) had
noted that the rich but low density
cluster A3574, which is very spiral rich (Willmer et al. 1991), 
had a very high ratio of low surface brightness (LSB) dwarfs to
giants. This is now backed up by the observations
of clusters
like A204 which are dwarf rich (DGR $\sim 3$), have low central densities 
and late B-M types (A204 is B-M III). 

\begin{figure}
\plotone{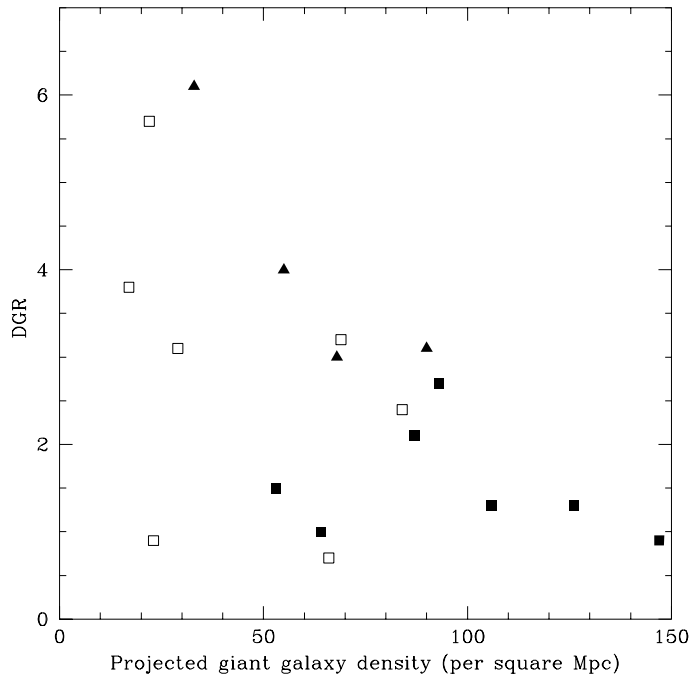}
\caption{Variation of the dwarf-to-giant ratio (DGR), as defined
in the text, with projected density of cluster giants (per square Mpc).
Solid boxes represent the central 1 square Mpc regions of the clusters,
the open boxes the outer regions (data from Driver at al 1998). 
The triangles show the variation over
a wider range of radii for Abell 2554 (data from Smith et al 1997). 
Note that typical
error bars (due to the combination of Poisson errors and background
subtraction errors) are 10\% in density and 20\% in DGR for the denser regions,
rising to 30\% in density and 50\% in DGR at the lowest densities (and hence
object numbers). The outlier at low density and low DGR (the outskirts
of A22) has a very large error in DGR ($\sim$ 100\%).
\label{fig1}}
\end{figure}

To extend the range of environments studied, we can add in further
LF results from the literature (Figure 2).
A problem here, of course, is the lack of homogeneity
due to different observed wavebands, different object detection
techniques and so forth.
Nevertheless, we can explore the general trends. 
Several points are shown for surveys of Coma (hexagons).
These surveys (Thompson \& Gregory 1993, Lobo et al. 1997, Secker \& Harris 
1996 and
Trentham 1998) cover different areas and hence different mean projected
densities (see also the next section). All these lie close to the relation
defined by our original data, with the larger area surveys having higher DGRs.
Points (filled triangles)
representing the rich B-M type I X-ray selected clusters studied by
Lopez-Cruz et al. (1997) fall at
somewhat lower DGR than most of our clusters at similar densities. However
we should note that these clusters were selected (from a larger unpublished 
sample) {\it only} if they had LFs well fitted by a single Schechter
function. This obviously precludes clusters with steep LF turn-ups
at intermediate magnitudes and hence rules out high DGRs. 
The one comparison cluster they do show {\it with} a
turn up (A1569 at DGR $\simeq 4.2$) clearly supports our overall trend. 
 
Ferguson \& Sandage (1991 = FS), on the other hand, deduced a trend in
the opposite direction, from a study of
fairly poor groups and clusters, with the early type dwarf-to-giant 
ratio {\it increasing} for
denser clusters. However, this is not necessarily as contradictory to the
present result as it might initially appear. For instance, FS select
their dwarfs morphologically, not by luminosity (morphologically
classified dwarfs and giants significantly overlap in luminosity) and
they also concentrate solely on early type dwarfs. If, as we might expect,
low density regions have significant numbers of 
late type dwarf irregulars (e.g. Thuan et al. 1991), then
the FS definition of DGR may give a lower value than ours for these regions.
Furthermore FS calculate their projected densities from {\it all} detected
galaxies, down to very faint dwarfs. Regions with high DGR will therefore
be forced to much higher densities than we would calculate for giants only.
These two effects may go much of the way to reconciling our respective
results. This is illustrated by the open triangles in Figure 2, which are
an attempt to place the FS points on our system; magnitudes have
been adjusted approximately for the different wavebands, 
DGRs have been estimated
from the LFs and the cluster central 
densities (from Ferguson \& Sandage 1990)
have been scaled down by the fraction of
their overall galaxy counts which are giants (by our luminosity definition).
Given the uncertainties in the translation, 
most of the FS points then lie  
close to those of our overall distribution.
Finally, a field LF with a steep faint end tail ($\alpha
\simeq -1.5$; e.g. Marzke, Huchra \& Geller 1994, Zucca et al. 1997, Morgan,
Smith \& Phillipps 1998) 
would also give a point (filled pentagon) at DGR $\simeq 4$, again
consistent with the trend seen in the clusters.

Nevertheless, there are exceptions. The FS points of lowest density 
(the Leo and Dorado groups) also have low DGR (and lie close to {\it our} 
main `outlier', the point for the outer region of A22). 
The Local Group (shown by the star)
would also be in this regime, at low density and DGR = 2,
as would the `conventional' field with $\alpha \simeq -1.1$ (Efstathiou, Ellis
\& Peterson 1988; Loveday et al. 1992) and hence DGR
$\simeq 1.5$ (open pentagon). 
This may suggest that at very low density 
the trend is reversed (i.e. is in the direction seen by FS), 
or that the cosmic (and/or statistical) scatter 
becomes large. More data in the very low density regime is probably
required before we can make a definitive statement on a possible reversal
of the slope of the DGR versus density relation. In particular, the scatter
in the derived faint end of the field LF between different surveys
(see, e.g., the recent discussion
in Metcalfe et al. 1998) precludes using this to tie
down the low density end of the plot.

\begin{figure}
\plotone{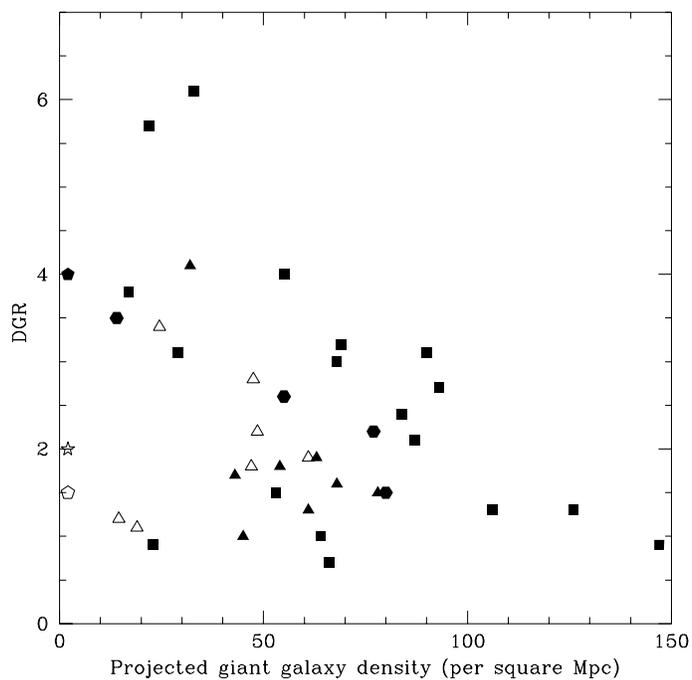}
\caption{
As Figure 1, but including data from other observers. Squares are our
data repeated from Figure 1, hexagons are for various Coma surveys detailed
in the text, filled triangles are from Lopez-Cruz's sample and open 
triangles are for Ferguson and Sandage's poor clusters and groups. 
The open pentagon at low density represents a conventional `flat' field LF, 
the filled pentagon a possible
steep ($\alpha \simeq -1.5$) field LF and the Local Group
is represented by the star at DGR = 2.
\label{fig2}}
\end{figure}

\subsection{Population Gradients}

It was suggested by the results on A2554 (Smith et al. 1997), that the dwarf
population was more spatially extended than that of the giants, i.e. the
dwarf to giant ratio increased outwards. This type of population gradient 
has now been confirmed by the results in Driver et al. (1998) illustrated
in Figure 1, where we contrast the inner 1 Mpc$^2$ areas (solid symbols)
with the outer regions of the same clusters (open symbols). The triangles show 
in slightly more detail the run of DGR with radius (and hence density)
across A2554. A similar effect can be seen for Coma in Figure 2
and can explain the discrepancy between the LFs derived for the core
and as against larger areas. 
It is found, too, in Virgo 
(Phillipps et al. 1998a; Jones et al., these proceedings), 
where the dwarf LSBG population has almost constant
number density across the central areas while the giant density drops
by a factor $\sim 3$.

\section{A Dwarf Population Density Relation}

The obvious synthesis of the above results is a relationship
between the {\it local} galaxy density and the fraction of dwarfs 
(i.e. the relative
amplitude of the dwarf LF). The inner, densest parts of rich clusters
have the smallest fraction of dwarfs, while loose clusters and the
outer parts of regular clusters, where the density is low, have high dwarf
fractions. It is particularly interesting
to note the clear overlap region in Figure 1, 
where regions of low density on the
outskirts of dense clusters (open squares) have similar DGRs to the
regions of the same density at the centres of looser clusters (solid squares).

The proposed relation
of course mimics the well known morphology - density relation (Dressler 
1980), wherein the central parts of rich clusters have the highest early type
galaxy fraction, this fraction then declining with decreasing local galaxy 
density.
Putting the two relations together, it would also imply that dwarfs 
preferentially occur in the same environments as spirals. This would be in 
agreement with the weaker clustering of low luminosity systems in general
(e.g. Loveday et al. 1995), as well as for spirals compared to ellipticals
(Geller \& Davies 1976). Thuan et al. (1991) have previously discussed the
similar spatial distributions of dwarfs (in particular dwarf irregulars)
and larger late type systems.

\section{The Origin of the Relation}

As with the corresponding morphology - density relation for giant galaxies, the
cause of our population - density relation could be either `nature' or
`nurture', i.e. initial conditions or evolution. Some clues may be provided by
the most recent semi-analytic models of galaxy formation, which have been 
able to account successfully for the excess of (giant) early type
galaxies in dense environments (e.g. Baugh, Cole \& Frenk 1996), basically
through different merging histories for different types of galaxy.
Does this also work for the dwarfs? 

The steep faint end slope of the LF appears to be a generic result of
hierarchical clustering models 
(e.g. White \& Frenk 1991; Frenk et al. 1996;
Kauffmann, Nusser \& Steinmetz 1997 = KNS),
so is naturally accounted
for in the current generation of models. The general hierarchical
formation picture envisages (mainly baryonic) galaxies forming at the cores
of dark matter halos. The halos themselves merge according to the general
Press-Schechter (1974) prescription to generate the present day halo mass
function. However the galaxies can retain their individual 
identities within the
growing dark halos, because of their much longer merging time scales.
The accretion of small halos by a large one then results in the main
galaxy (or cluster of galaxies, for very large mass halos) acquiring a
number of smaller satellites (or the cluster gaining additional, less 
tightly bound, members).

KNS have presented a detailed study of the distribution
of the luminosities of galaxies expected to be associated with a single
halo of given mass.  We can thus easily compare the theoretically
expected numbers of dwarf galaxies
per unit giant galaxy luminosity with our empirical results (Phillipps et al.
1998b). 

The KNS models mimic a ``Milky Way system" 
(halo mass $5 \times 10^{12} M_{\odot}$), a sizeable group 
(halo mass $5 \times 10^{13} M_{\odot}$)
and a cluster mass halo ($10^{15} M_{\odot}$).
Their results imply
that the Milky Way and small group halos have similar numbers
of dwarf galaxies per unit giant galaxy light, whereas the dense cluster 
environment has a much smaller number of dwarfs for a given total giant
galaxy luminosity. Thus the predictions of the hierarchical models
(which depend, of course, on the merger history of the galaxies)
are in qualitative agreement with our empirical results if we identify
loose clusters and the outskirts of rich clusters with a population
of (infalling?) groups (cf. Abraham et al. 1996), 
whereas the central dense regions of the clusters 
originate from already massive dark halos. 
If we renormalise from unit galaxy light to an
effective giant galaxy LF amplitude (see Phillipps et al. 1998b)
then the actual expected ratios
($\sim 1$ to a few) are also consistent with our observational results.

By inputting realistic
star formation laws etc., KNS could further identify the
galaxies in the most massive halos with old elliptical galaxies, and
those in low mass halos with galaxies with continued star formation.
This would imply the likelihood that our dwarfs in low density regions
may still be star forming, or at least have had star formation in the
relatively recent past (cf. Phillipps \& Driver 1995 and references 
therein). Note, too, that these galaxy formation models would also 
indicate that the usual (giant) morphology - density relation and our 
(dwarf) population - density relation {\it do} arise in basically the same way.
Finally, we can see that if these semi-analytic
models are reasonably believable,
then we need not necessarily expect the field to be even richer in dwarfs than
loose clusters; the dwarf to giant ratio seems to level off at the
densities reached in fairly large groups.

\section{Summary}

To summarise, then,
we suggest that the current data on the relative numbers of dwarf
galaxies in different clusters and groups can be understood in terms
of a general dwarf population versus local galaxy density relation,
similar to the well known morphology - density relation for giants.
Low density environments are the preferred habitat of low luminosity
galaxies; in dense regions they occur in similar numbers to giants,
but at low densities dwarfs dominate numerically by a large factor. 
This fits in with the general idea that low luminosity
galaxies are less clustered than high luminosity ones (particularly
giant ellipticals). Plausible theoretical justifications for the 
population - density relation can be found within the context of current
semi-analytic models of hierarchical structure formation.


\begin{references}
\reference Abraham R.G., et al., 1996, ApJS, 471,  694
\reference Baugh C.M., Cole S., Frenk C.S., 1996, MNRAS, 283, 1361
\reference Dressler A., 1980, ApJ, 236, 351
\reference Driver S.P., Phillipps S., Davies J.I., Morgan I., 
    Disney M.J., 1994, MNRAS, 268, 393 
\reference Driver S.P., Couch W.J., Phillipps S., 1998, MNRAS, 301, 369
\reference Efstathiou G., Ellis R.S., Peterson B.A., 1988, MNRAS, 232, 431
\reference Ferguson H.C., Sandage A., 1990, AJ, 100, 1
\reference Ferguson H.C., Sandage A., 1991, AJ, 96, 1520
\reference Frenk C.S., Evrard A.E., White S.D.M., Summers F.J., 1996,
    ApJ, 472, 460
\reference Geller M.J., Davis M., 1976, ApJ, 208, 13
\reference Jones J.B., Phillipps S., Schwartzenberg J.M., Parker Q.A., 1998,
The Low Surface Brightness Universe, in press
\reference Kauffmann G., Nusser A., Steinmetz M., 1997, MNRAS, 286, 795
\reference Lobo C., et al., 1997, A\&A, 317, 385
\reference Lopez-Cruz O., Yee H.K.C., Brown J.P., Jones C., Forman W., 
    1997, ApJL, 475, L97
\reference Loveday J., Maddox S.J., Efstathiou G., Peterson B.A., 1995, 
    ApJ, 442, 457
\reference Loveday J., Peterson B.A., Efstathiou G., Maddox S.J., 1992, 
    ApJ, 390, 338
\reference Marzke R., Huchra J.P., Geller M.J., 1994, ApJ, 428, 43
\reference Metcalfe N., Ratcliffe A., Shanks T., Fong R., 1998, 
    MNRAS, 294, 147
\reference Morgan I., Smith R.M., Phillipps S., 1998, MNRAS, 295, 99
\reference Phillipps S., Driver S.P., 1995, MNRAS, 274, 832
\reference Phillipps S., Driver S.P., Couch W.J., Smith R.M., 
    1998b, ApJ, 498, L119
\reference Phillipps S., Parker Q.A., Schwartzenberg J.M., 
    Jones J.B., 1998a, ApJ, 493, L59
\reference Press W.H., Schechter P.L., 1974, ApJ, 187, 425
\reference Schechter P., 1976, ApJ, 203, 297
\reference Secker J., Harris W.E., 1996, ApJ, 469, 623
\reference Smith R.M., Driver S.P., Phillipps S., 1997, MNRAS, 287, 415
\reference Smith R.M., Phillipps S., Driver S.P., Couch R.M., 1998, The
    Low Surface Brightness Universe, in press
\reference Thompson L.A., Gregory S.A., 1993, AJ, 106, 2197
\reference Thuan T.X., Alimi J.M., Gott J.R., Schneider S.E., 1991, 
    ApJ, 370, 25
\reference Trentham N., 1997a, MNRAS, 286, 133
\reference Trentham N., 1997b, MNRAS, 290, 334
\reference Trentham N., 1998, MNRAS, 293, 71
\reference Turner J.A., Phillipps S., Davies J.I., Disney M.J., 1993, 
    MNRAS, 261, 39
\reference White S.D.M., Frenk C.S., 1991, ApJ, 379, 52
\reference Willmer C., Focardi P., Chan R., Pellegrini P., 
    da Costa L., 1991, AJ, 101, 57
\reference Wilson G, Smail I., Ellis R.S., Couch W.J., 1997, 
    MNRAS, 284, 915
\reference Zucca E., et al., 1997, in Wide-Field Spectroscopy, 
    eds. Kontizas E. et al., Dordrecht; Reidel, p.247
\end{references}
\end{document}